\documentclass[12pt]{article}
\usepackage{amsfonts,amsmath,amsthm}
\usepackage{graphicx}
\usepackage{rotating}

\def\be{\begin{equation}}
\def\ee{\end{equation}} 
\def\ba{\begin{array}}
\def\ea{\end{array}}
\def\cR{{\cal R}}

\begin{document}
\title {Quantifying the transmission potential of pandemic influenza}
\author{Gerardo Chowell$^{1}$ \thanks{Corresponding author. Email: gchowell@asu.edu Fax: (505) 480-965-7671}, Hiroshi Nishiura$^{2}$\\
\footnotesize $^{1}$ School of Human Evolution and Social Change, Arizona State University\\
\footnotesize Tempe, AZ 85282, USA\\
\footnotesize $^{2}$ Theoretical Epidemiology, University of Utrecht,\\ 
\footnotesize   Yalelaan 7, 3584 CL Utrecht, The Netherlands \\
}

\date{}

\maketitle

\newpage 

\begin{abstract}
This article reviews quantitative methods to estimate the basic reproduction number of pandemic influenza, a key threshold quantity to help determine the intensity of interventions required to control the disease.  Although it is difficult to assess the transmission potential of a probable future pandemic, historical epidemiologic data is readily available from previous pandemics, and as a reference quantity for future pandemic planning, mathematical and statistical analyses of historical data are crucial. In particular, because many historical records tend to document only the temporal distribution of cases or deaths (i.e. epidemic curve), our review focuses on methods to maximize the utility of time-evolution data and to clarify the detailed mechanisms of the spread of influenza. 
\\
\indent First, we highlight structured epidemic models and their parameter estimation method which can quantify the detailed disease dynamics including those we cannot observe directly. Duration-structured epidemic systems are subsequently presented, offering firm understanding of the definition of the basic and effective reproduction numbers. When the initial growth phase of an epidemic is investigated, the distribution of the generation time is key statistical information to appropriately estimate the transmission potential using the intrinsic growth rate. Applications of stochastic processes are also highlighted to estimate the transmission potential using similar data. Critically important characteristics of influenza data are subsequently summarized, followed by our conclusions to suggest potential future methodological improvements.
\end{abstract}

\newpage 

\textbf{PACS classifications:}  Viral diseases (87.19.xd); Population dynamics (87.23.Cc); Stochastic models in biological physics (87.10.Mn)

\textbf{Keywords:}  Influenza; pandemic; epidemiology; basic reproduction number; model.

\tableofcontents

\section{Introduction}
\linespread{1.9}

Influenza epidemics are observed around the world during the wintertime and with a strong seasonal component in temperate regions \cite{Simonsen1}\cite{Park2007}. Influenza is a disease caused by the influenza virus, an RNA virus belonging to the Orthomyxoviridae \cite{Webster1}. Many features are common with those of the paramyxovirus infections of the acute upper respiratory tract. Typical symptoms of the disease are characterized by fever, myalgia, severe malaise, non-productive cough, and sore throats. The disease spreads when an infected individual coughs or sneezes and sends the virus into the air, and other susceptible individuals inhale the virus. The virus is also believed to be transmitted when a person touches a surface that is contaminated with the virus ({\it e.g.} door knob, etc.) and then touches the nose or eyes. Infected individuals can transmit the virus almost within a day following infection ({\it i.e.} latent period). Although it is generally believed that infected individuals can pass the virus for 3-7 days following symptom onset, there is some uncertainty on the duration of the infectious period. The generation time ({\it i.e.} sum of latent and infectious periods) for influenza, reported and assumed in the literature, ranges from 3 days \cite{Ferguson1}\cite{Ferguson2}\cite{Wallinga2} to about 6 days \cite{Longini1}\cite{Mills1}.
 
Individuals that are infected with influenza are believed to become permanently immune against the specific virus strain. Hence, the virus is able to persist in the human population through relatively minor (single point) mutations in the virus composition known as drifts. Influenza (sub)types A/H3N2, A/H1N1 and B are currently co-circulating in the human population \cite{Nicholson1}. Major changes in the virus composition via recombination or gene reassortment processes (known as genetic shifts) can lead to the emergence of novel influenza viruses with the potential of generating dramatic morbidity and mortality levels around the world \cite{Webster1}.
 
The 1918-19 influenza pandemic known as the {\bf Spanish influenza} caused by the influenza virus A (H1N1) has been the most devastating in recent history with estimated worldwide mortality ranging from 20 to 100 million deaths \cite{Cunha1}\cite{Murray2006} with a case fatality of $2$-$6$ percent \cite{Sydenstricker1921}\cite{Markel2007}. The worldwide 1918 influenza pandemic spread in three waves starting from Midwestern United States in the spring of 1918  \cite{Patterson1}\cite{Johnson1}. The deadly second wave began in late August probably in France while the third wave is generally considered as part of normal more scattered winter outbreaks similar to those observed after the 1889/90 pandemic \cite{Patterson1}. Subsequent pandemics during the 20th century are attributed to subtyes A (H2N2) from 1957-58 (Asian influenza) and A (H3N2) in 1968 (Hong Kong influenza) \cite{MacKellar2007}. 

The ability to quickly detect and institute control efforts at the early stage of an influenza pandemic is directly linked to the final levels of morbidity and mortality in the population \cite{Markel2007}. To appropriately assess the disaster size of a probable future pandemic, we have to quantify the transmission potential (and its associated uncertainty). Although it is difficult to directly measure the transmissibility of a future pandemic, historical epidemiologic data is readily available from previous pandemics, and as a reference quantity for future pandemic planning, mathematical and statistical analyses of the historical data can offer various insights. In particular, because many historical records tend to document only the temporal distribution of cases or deaths ({\it i.e.} epidemic curve), we modelers have faced with a difficult need to clarify the mechansms of the spread of influenza using such time-evolution data alone. In this paper, we review a number of mathematical and statistical methods for the estimation of the transmission potential of pandemic influenza, focusing on theoretical techniques to maximize the utility of the temporal distribution of influenza cases. The methods that have been incorporated in this review include the applications of epidemiologically structured epidemic models, explicitly duration-structured epidemic system, and stochastic processes ({\it i.e.} branching and counting processes). Whereas this review does not cover the spread of influenza in space, spatial heterogeneity in transmission and the growing interest in the role of contact network are briefly discussed as the future challenge.

\section{On the definition of the transmission potential}

The basic reproduction number, $R_0$ (pronounced as {\it R nought}), is a key quantity used to estimate transmissibility of infectious diseases. Theoretically, $R_0$ is defined as the average number of secondary cases generated by a single primary case during its entire period of infectiousness in a completely susceptible population \cite{Diekmann2000book}. As the epidemic progresses, the number of susceptible individuals is decreased due to infection, and the reproduction number decays following $R(t) = R_0S(t)/S(0)$ where $S(t)$ and $S(0)$ are, respectively, the number of susceptible individuals at time $t$ and before the epidemic starts; the latter is equivalent to the total population size $N$ given that all individuals are susceptible before the beginning of an epidemic. Clearly, this definition only applies to (novel) emerging infectious diseases ({\it e.g.} the epidemic of severe acute respiratory syndrome (SARS) from 2002-3) or re-emerging infectious diseases that had not circulated in the population in question for long enough to allow for residual immunity in the population to disappear due to births and deaths. 
\\
\indent The reproduction number is directly related to the type and intensity of interventions necessary to control an epidemic since the objective is to make $R(t)<1$ as soon as possible. To achieve $R(t)<1$, one or a combination of control strategies may be implemented. For example, one of the best known uses of $R_0$ is in determining the critical coverage of immunization required to eradicate a disease in a randomly mixing population. That is, when vaccine is available against a disease in question, it is of interest to estimate the critical proportion of the population that needs to be vaccinated ({\it i.e.} vaccination coverage) in order to attain $R<1$ \cite{Anderson1991}\cite{Anderson1982}. For example, in the U.S prior to 1963, a vaccine against measles was not available and hence recurrent epidemics of measles were observed with approximately $3-4$ million cases per year and a mean of $450$ deaths. The introduction of the vaccine in the U.S. reduced the incidence by 98 percent.
\\
\indent The critical vaccination coverage, $p_c$ (in a randomly mixing population) can be estimated from the $R_0$ of the disease in question as follows \cite{Smith1964}:\\
\begin{equation}
\label{eqn_Intro1}
p_c>\dfrac{1}{\epsilon}\left(1-\dfrac{1}{R_0}\right)
\end{equation}
where $\epsilon$ is the efficacy ({\it i.e.} direct effectiveness) of vaccination \cite{Halloran1991}. $p_c$ given in (\ref{eqn_Intro1}) suggests that the disease could be eradicated even when all susceptible individuals are not vaccinated. The protection conferred to the population by achieving a critical vaccination coverage is known as {\bf herd immunity} \cite{Fine1993}\cite{Nishiura2006Lotz}. 
\\
\indent A brief history of the theoretical developments on the basic reproduction number and its analytical computation via epidemic modeling is given elsewhere \cite{Nishiura2006Lotz}\cite{Heesterbeek1}\cite{Nishiura2007R0}. The mathematical definition and calculation of $R_0$ using next-generation arguments is described initially by Odo Diekmann and his colleagues \cite{Diekmann2000book}\cite{Diekmann2}, where $R_0$ is the dominant eigenvalue of the resulting next generation matrix. Further elaborations and reviews can be found elsewhere \cite{vanden1}\cite{Carlos1}\cite{Hyman1}\cite{Heffernan2005}\cite{Heffernan2006}\cite{Nishiura2006den}\cite{Keeling2000R0}. Classically, rather than the threshold phenomena, $R_0$ was used to suggest the {\it severity} of an epidemic, because the proportion of those experiencing infection at the end of an epidemic depends only on $R_0$ \cite{Kendall1956} (see Section 3).
\\
\indent Statistical methods to quantitatively estimate $R_0$ have been reviewed by Klaus Dietz \cite{Dietz1993}. Depending on the characteristics of data and underlying assumptions of the models, $R_0$ can be estimated using various different approaches \cite{DeJong1994}. In addition to the final size equation, $R_0$ of an epidemic of newly emerging disease can be estimated from the {\bf intrinsic growth rate} \cite{Ferguson1}\cite{Wallinga2}\cite{Mills1}\cite{Anderson1991}\cite{Lipsitch1}\cite{Chowell1}, which is also referred to as the {\it rate of natural increase}, suggesting the natural growth rate of infected individuals in a fully susceptible population (discussed in Section 4). Moreover, for simple epidemic models with relatively few parameters, $R_0$ can be estimated  with other unobservable quantities by rigorous curve fitting of model equations to the observed epidemic data (discussed in Section 3)\cite{Chowell1}\cite{Chowell2}\cite{ChowellMBE}. Not only $R_0$ but also $R(t)$ can be estimated from the temporal distribution of infectious diseases, reconstructing the transmission network or inferring the time-inhomogeneous number of secondary transmissions \cite{Wallinga1}\cite{Cauchemez1}\cite{Bettencourt1}\cite{Nishiura2006}.
\\
\indent To estimate the basic reproduction number of endemic diseases, different approaches are taken. One would need first to carry out serological surveys to quantify the fraction of the population that is effectively protected against infection ({\it i.e.} age- and/or time-specific proportion of those possessing acquired immunity needs to be estimated). Through this effort, the {\bf force of infection}, the rate at which susceptible individuals get infected, is estimated \cite{Muench1959}. For example, this is the case for rubella, mumps and measles (that are still circulating in some regions of the world even when high effective vaccination coverage is achieved). Although the estimation of $R_0$ for endemic diseases is out of the scope of this review, methodological details and the applications to estimate the force of infection and $R_0$ can be found elsewhere \cite{Schenzle1979}\cite{Farrington1990}\cite{Keiding1991}\cite{Ades1993}\cite{Whitaker2004}\cite{Farrington2001R0}\cite{Satou2007HEV}.
\\
\indent In practice, the reproduction number denoted simply by $R$ and defined as the number of secondary cases generated by a primary infectious cases in a partially protected population might be useful. $R$ can also be estimated from the initial growth phase of an epidemic in such a partially immunized population. In a randomly mixing population, the relationship between the basic reproduction number ($R_0$) and the reproduction number ($R$) is given by  $R=(1-p)R_0$ where $p$ is the proportion of the population that is effectively protected against infection (in the beginning of an epidemic). Besides, for many recurrent infectious diseases including seasonal influenza, estimating the background immunity $p$ in the population is extremely difficult due to cross-immunity of antigenically-related influenza strains and vaccination campaigns. 
\\
\indent With reagard to seasonal influenza, the reproduction number ($R$) over the last 3 decades has been estimated at $1.3$ (SE $0.05$) in the United States, France, and Australia with an overall range of $0.9-2.0$ \cite{Chowell3}. An $R$ estimate of $1.5$ has been reported for a single A/H3N2 season in France \cite{Flahault1}, and some estimates have been reported in the range $1.4$-$2.6$ for several consecutive influenza seasons in England and Wales \cite{Spicer1}\cite{Spicer2}. A particularly high estimate of $R$ has been suggested for the 1951 influenza epidemic in England and Canada \cite{Viboud2}.
\\
\indent Because influenza pandemics such as the Spanish flu from 1918-19 are associated to the emergence of novel influenza strains to which most of the population is susceptible, it might be reasonable to assume that the reproduction number $R\approx R_0$. Previous studies have estimated that $R_0$ of the 1918-19 influenza pandemic ranged between $1.5$ and $5.4$ \cite{Mills1}\cite{Chowell1}\cite{ChowellMBE}\cite{Chowell2}\cite{Viboud2}\cite{Gani1}\cite{Massad1}\cite{Nishiura1}\cite{Sertsou1}\cite{Andreasen1}\cite{Vynnycky2007}\cite{Mathews1} depending on the specific location and pandemic wave considered, type of data, estimation method, and level of spatial aggregation, which has ranged from small towns to entire nations with several million inhabitants. Table \ref{TableWaves} lists estimates of $R_0$ in recent studies. The variability of $R_0$ estimates suggests that local factors, including geographic and demographic conditions, could play an important role in disease spread \cite{ChowellRev}\cite{Sattenspiel1998}\cite{Sattenspiel2003}. In the following sections, we review how these estimates are obtained and how we shall interpret the estimates, starting from a simple structured epidemic model proposed in 1927.

\section {Estimating $R_0$ using a structured epidemic model}
\begin{figure}
   \begin{center}
   {\includegraphics[width=15cm]{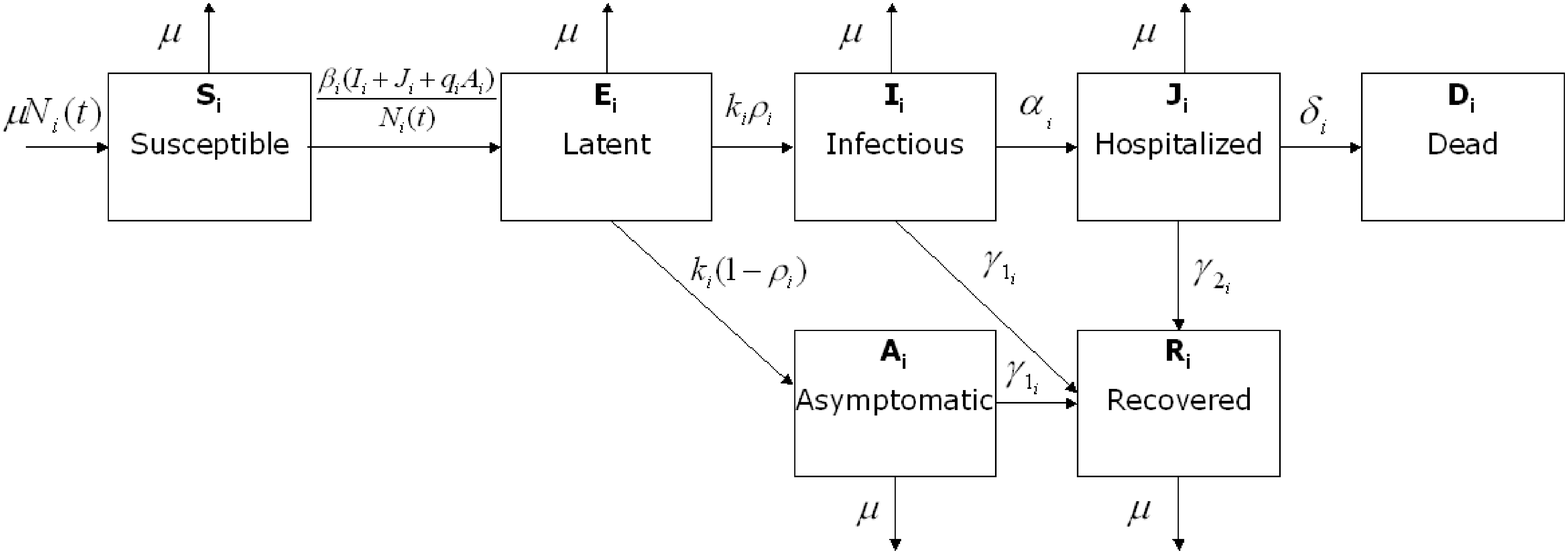}}
   \end{center}
   \caption{{\bf Flow chart of the state progression of individuals among the different epidemiological classes as modeled by the complex SEIR model.} See equations (\ref{eqn2}).}
\label{figDiagram}
\end {figure}

Mathematical models provide a unique way to analyze the transmission dynamics and study various different scenarios associated to the spread of communicable diseases in population(s) \cite{Anderson1991}\cite{BC}\cite{Hethcote1}. The history of the mathematical modeling of infectious diseases greatly remounts to the study of Sir Ronald Ross in 1911 \cite{Ross1} who invented a classic malaria model and also discovered the mosquito-borne transmission mechanisms of malaria. Employing a mass action principle for the spread of malaria, Ross explored the effects of controling the mosquito population using simple mathematical models \cite{Dietz1988}. Following his effort, Kermack and Mckendrick introduced a classical SIR (susceptible-infectious-removed) epidemic model in 1927, which is most frequently utilized in the present day, given by the following system of nonlinear ordinary differential equations (ODEs) \cite{Kermack1927}:

\begin{equation}
\label{eq1}
\begin{array}{rcl}
  \dfrac{dS(t)}{dt} &=& -\dfrac{\beta  S(t) I(t)}{N} \\
  \dfrac{dI(t)}{dt} &=& \dfrac{\beta  S(t) I(t)}{N} -\gamma I(t) \\
  \dfrac{dR(t)}{dt} &=& \gamma I(t) \\
\end{array}
\end{equation}

\noindent where $S(t)$ denotes susceptible individuals at time $t$; $I(t)$, infected (assumed infectious) individuals at time $t$; and $R(t)$, recovered (assumed permanently immune) individuals at time $t$; $\beta$ is the transmission rate; $\gamma$ the recovery rate; and $N$ is the total population size which is assumed constant for a closed population ({\it i.e.} a population without immigration and emmigration). Susceptible individuals in contact with the virus enter the infectious class ($I$) at the rate $\beta I/N$. That is, homogeneous mixing between individuals is assumed.\\
\indent The basic reproduction number, $R_0$, for the epidemic system (\ref{eq1}) is given by the product of the transmission rate and the mean infectious period. That is:
\begin{equation}
\label{eq_R0}
R_0 = \dfrac{\beta}{\gamma}
\end{equation}
\noindent Classically, $R_0$ has been known as a quantity to suggest {\bf severity} of an epidemic \cite{Kendall1956}. Indeed, analytical expression of $R_0$ in (\ref{eq_R0}) is derived simply by solving the above system (\ref{eq1}). Replacing $I(t)$ in the right hand side of $dS(t)/dt$ by ($1/\gamma$)$dR(t)/dt$, we get
\begin{equation}
\label{fsize1}
  \dfrac{1}{S(t)} \dfrac{dS(t)}{dt} = -\dfrac{\beta }{\gamma N} \dfrac{dR(t)}{dt}
\end{equation}
\noindent Integrating both sizes of (\ref{fsize1}) from 0 to infinity,
\begin{equation}
\label{fsize2}
  \ln  \dfrac{S(\infty)}{S(0)} = -\dfrac{\beta }{\gamma N} (R(\infty)-R(0))
\end{equation}
Since $S(\infty)=N-R(\infty)$, and because we assume $S(0)=N$ and $R(0)=0$, equation (\ref{fsize2}) can be rewritten as
\begin{equation}
\label{fsize3}
  \ln  \dfrac{N-R(\infty)}{N} = -\dfrac{\beta }{\gamma} \dfrac{R(\infty)}{N}
\end{equation}
In the above equation (\ref{fsize3}), {\bf final size}, {\it i.e.}, the proportion of those experiencing infection among a total number of individuals in a community following a large scale epidemic, is defined as $p:=R(\infty)/N$. That is,
\begin{equation}
\label{fsize4}
  \ln  (1-p) = -R_0 p
\end{equation}
Therefore, the following {\bf final size equation} of an autonomous SIR (or SEIR) model is obtained:
\begin{equation}
\label{fsize5}
\hat{R_0} = \dfrac{-\ln(1-p)}{p}
\end{equation}
Equation (\ref{fsize5}) can be analytically derived using both deterministic (models governed by ODEs or partial differential equations (PDEs))\cite{Ma2006} and stochastic models \cite{Becker1989}.
\indent Despite the usefulness of (\ref{eq1}), SIR assumptions given by ODEs are not always directly applicable to real data. One of the reasons include that there is no disease where an infected individual can cause secondary transmission immediately after his/her infection. 
\\
\indent Accordingly, we have used slightly extended compartmental models in the previous studies to describe the transmission dynamics of the 1918-19 influenza pandemic and estimate the reproduction number \cite{Chowell1}\cite{Chowell2}.  We now describe two different SEIR (susceptible-exposed-infectious-removed) models that have been used to estimate the reproduction number. The first model is the simple SEIR model, and the second model accounts for asymptomatic and hospitalized individuals. 
\\
\indent The simple SEIR model classifies individuals as susceptible (S), exposed (E), infectious (I), recovered (R), and dead (D) \cite{Anderson1991}. Susceptible individuals in contact with the virus enter the exposed class at the rate $\beta I(t)/N$, where $\beta$ is the transmission rate, $I(t)$ is the number of infectious individuals at time $t$ and $N = S(t) + E(t) + I(t) + R(t)$ is the total population for any $t$. The entire population is assumed to be susceptible at the beginning of the epidemic. Individuals in latent period (E) progress to the infectious class at the rate $k$  (where $1/k$ suggests the mean latent period). We assume homogeneous mixing ({\it i.e.} random mixing) between individuals and, therefore, the fraction $I(t)/N$ is the probability of a random contact with an infectious individual in a population of size $N$. Since we assume that the time-scale of the epidemic is much faster than characteristic times for demographic processes (natural birth and death), background demographic processes are not included in the model. Infectious individuals either recover or die from influenza at the mean rates $\gamma$ and $\delta$, respectively. Recovered individuals are assumed protected for the duration of the outbreak. The mortality rate is given by $\delta = \gamma$ [CFP/(1-CFP)], where CFP is the mean case fatality proportion. The transmission process can be modeled using the system of nonlinear differential equations:
\begin{equation}
\label{eqn12}
\left \{ \begin{array}{rcl}
      \dfrac{dS(t)}{dt} &=& -\dfrac{\beta S(t) I(t)}{N}\\
      \dfrac{dE(t)}{dt} &=& \dfrac{\beta S(t) I(t)}{N(t)} - k  E(t)\\
      \dfrac{dI(t)}{dt} &=& k  E(t) -(\gamma + \delta) I(t)\\ 
      \dfrac{dR(t)}{dt} &=& \gamma I(t)\\
     \dfrac{dD(t)}{dt} &=& \delta I(t)\\
     \dfrac{dC(t)}{dt} &=& k  E(t)\\
  \end{array} \right.
\end{equation}
where $C(t)$ is the cumulative number of infectious individuals. The basic reproduction number of the above system (\ref{eqn12}) is given by the product of the mean transmission rate and the mean infectious period, $R_0 = \beta/(\gamma+\delta)$. \\
\indent A more complex SEIR model (Figure \ref{figDiagram}) classifies individuals as susceptible ($S$), exposed ($E$), clinically ill and infectious  ($I$), asymptomatic and partially infectious ($A$), diagnosed and reported ($J$), recovered ($R$), and death ($D$). The birth and natural death rates are assumed to have a common rate $\mu$ (60-year life expectancy as in \cite{Chowell1}). The entire population is assumed susceptible at the beginning of the pandemic wave. Susceptible individuals in contact with the virus progress to the latent class at the rate $\beta (I(t)+ J(t) + q A(t)) / N$ where $\beta$ is the transmission rate, and $0< q <1$ is a reduction factor in the transmissibility of the asymptomatic class ($A$). Since there is no explicit evidence estimating and proving the effectiveness of public health interventions, and because a high burden was placed upon the sanitary and medical sectors, diagnosed/hospitalized individuals ($J$) are assumed equally infectious. Although it is difficult to explicitly evaluate the difference in infectiousness between general community and hospital, we roughly made this assumption since 78 percent of the nurses of the San Francisco Hospital contracted influenza \cite{Hrenoff1}. A more rigorous assumption requires either statistical analysis of more detailed time-series data \cite{Cooper1} or an epidemiological comparison of specific groups by contact frequency \cite{Nishiura2}. The total population size at time $t$ is given by $N=S(t)+E(t)+I(t) + A(t) + J(t) + R(t)$. We assumed homogeneous mixing of the population and, therefore, the fraction $(I(t) + J(t) + q A(t))/N$ is the probability of a random contact with an infectious individual. A proportion $0<\rho<1$ of latent individuals progress to the clinically infectious class ($I$) at the rate $k$ while the remaining ($1-\rho$) progress to the asymptomatic partially infectious class ($A$) at the same rate $k$ (fixed to 1/1.9 days$^{-1}$ \cite{Mills1}). Asymptomatic cases progress to the recovered class at the rate $\gamma_{1}$. Clinically infectious individuals (class $I$) are diagnosed (reported) at the rate $\alpha$ or recover without being diagnosed (e.g., mild infections, hospital refusals) at the rate $\gamma_{1}$. Diagnosed individuals recover at the rate $\gamma_{2} = 1/(1/\gamma_{1} - 1/\alpha)$ or die at rate $\delta$. The mortality rates were adjusted according to the case fatality proportion (CFP), such that $\delta = \dfrac{CFP}{1-CFP} (\mu + \gamma_{2})$.\\
\indent The transmission process can be modeled using the following system of nonlinear differential equations:
\begin{equation}
\label{eqn2}
\left \{ \begin{array}{rcl}
      \dfrac{dS(t)}{dt} &=& \mu N(t) -\dfrac{\beta S(t) (I(t) + J(t) + q A(t))}{N} - \mu S(t)\\
      \dfrac{dE(t)}{dt} &=& \dfrac{\beta S(t) (I(t) + J(t) + q A(t))}{N} -(k+\mu) E(t)\\ 
      \dfrac{dA(t)}{dt} &=& k (1-\rho) E  -(\gamma_{1} + \mu) A(t) \\ 
      \dfrac{dI(t)}{dt} &=& k \rho E(t) - (\alpha + \gamma_{1}  + \mu) I(t)\\
      \dfrac{dJ(t)}{dt} &=& \alpha I(t) - (\gamma_{2} + \delta + \mu) J(t)\\
      \dfrac{dR(t)}{dt} &=& \gamma_{1} (A(t) + I(t)) + \gamma_{2} J(t) -\mu R(t) \\
      \dfrac{dD(t)}{dt} &=& \delta J(t)\\
      \dfrac{dC(t)}{dt} &=& \alpha I(t)\\
  \end{array} \right.
\end{equation}
We assume the cumulative number of influenza notifications, our observed epidemic data, is given by $C(t)$. Seven model parameters ($\beta$, $\gamma_1$, $\alpha$, $q$, $\rho$, $E(0)$, $I(0)$) are estimated from the epidemic curve by least squares fitting as explained below.  The reproduction number for model (\ref{eqn2}) is given by (see \cite{Chowell1}):
\begin{equation}
\label{eqnR02}
\cR = \frac{\beta  k}{k+\mu} \Bigg \{ \rho \bigg( \frac{1}{\gamma_{1} + \alpha + \mu} + \frac{\alpha}{(\gamma_{1} + \alpha + \mu)(\gamma_{2} + \delta + \mu)} \bigg)  + (1-\rho) \bigg( \frac{q}{\gamma_{1} + \mu} \bigg) \Bigg \}
\end{equation}
and the clinical reporting proportion is given by:
\begin{equation}
\label{eqnO}
O = \frac{\alpha}{\alpha + \gamma_{1} + \mu}.
\end{equation}

\subsection{Parameter estimation}
In the simplest manner, model parameters can be estimated via least-square fitting of the model solution to the observed data. That is, one looks for the set of parameters $\bf{\hat{\Theta}}$ whose model solution best fits the epidemic data by minimizing the sum of the squared differences between the observed data $y_t$ and the model solution $C(t,\bf{\Theta})$. That is, we minimize:
\begin{equation}
\label{eqn20}
X(\bf{\Theta}) = \sum^n_{t=1} (y_t - C(t,\bf{\Theta}))^2
\end{equation} 
The standard deviation of the parameters can be estimated by computing the asymptotic variance-covariance $AV(\hat{\Theta})$ matrix of the least-squares estimate by \cite{Davidian1}:
\begin{equation}
\label{eqn21}
\bf{AV(\hat{\Theta})}= \sigma^2 (\nabla_{\Theta} \bf{C(\Theta_0)} \ \nabla_{\Theta} \bf{C(\Theta_0)^T})^{-1}
\end{equation} 
which can be estimated by
\begin{equation}
\label{eqn22}
\hat{\sigma}^2 (\hat{\nabla_{\hat{\Theta}}} \bf{C(\hat{\Theta})} \hat{\nabla_{\hat{\Theta}}} \bf{C(\hat{\Theta})^T})^{-1}
\end{equation} 
where $n$ is the total number of observations, $\hat{\sigma}^2$ is the estimated variance, and $\hat{\nabla} C$ are numerical derivatives of $C$. Estimates of $\hat{R_0}$ can be obtained by substituting the corresponding individual parameter estimates into an analytical formula of $R_0$.  Further, using the delta method \cite{Bickel1}, we can derive an expression for the variance of the estimated basic reproduction number $\hat{R_0}$. An expression for the variance of $R_0$ for the simple SEIR model (Equations \ref{eqn12}) is given by:

\begin{eqnarray} \label{varianceR0}
V(\hat{R_0})  & \approx  & \hat{R_0}^2 \ \{ \frac{V(\hat{\beta})}{\hat{\beta}^2} +
  \frac{V(\hat{\gamma})}{(\hat{\gamma} + \hat{\delta})^2}  + 
  \frac{V(\hat{\delta})}{(\hat{\gamma} + \hat{\delta})^2} \nonumber \\
                       &  &  - (\frac{2}{\hat{\beta} (\hat{\gamma} + \hat{\delta})}) (  Cov(\hat{\gamma}, \hat{\beta}) -  \frac{\hat{\beta} Cov(\hat{\delta}, \hat{\gamma})}{\hat{\gamma} + \hat{\delta}} +  Cov(\hat{\delta}, \hat{\beta}))  \}. 
\end{eqnarray}    

This expression depends on the variance (denoted by $V$) of the individual parameter estimates as well as their covariance (denoted by $Cov$).

\subsection{Bootstrap confidence intervals}
\begin{figure}[hp*]
   \begin{center}
   \scalebox{0.6}{\includegraphics{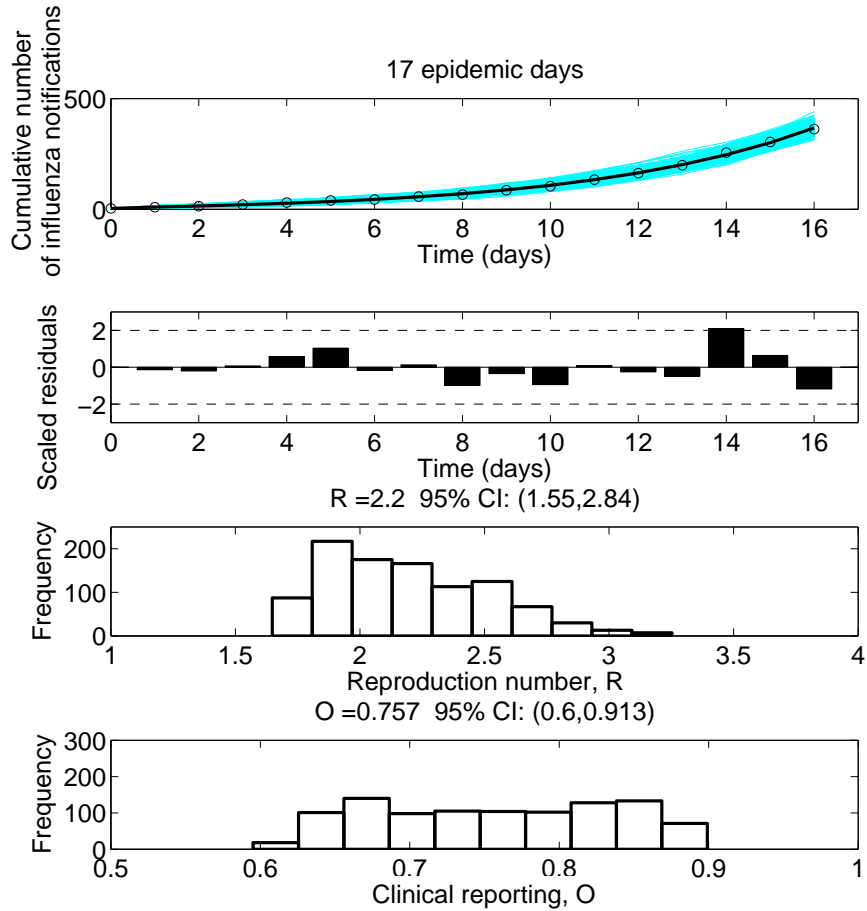}}
   \end{center}
   \caption{{\bf Model fits, residuals plots, and the resulting distributions of the reproduction number and the proportion of the clinical reporting obtained after fitting the complex SEIR epidemic model to the initial phase of the Fall influenza wave using 17 epidemic days of the Spanish Flu Pandemic in San Francisco, California}. See equations (\ref{eqn2}) \cite{Chowell2}. In the top panel, the epidemic data of the cumulative number of reported influenza cases are the circles, the solid line is the model best fit, and the solid blue lines are 1000 realizations of the model fit to the data obtained through parametric bootstrapping as explained in the main text.}
\label{figR-SEIJR}
\end {figure}

Another method to generate uncertainty bounds on the reproduction number is generating bootstrap confidence intervals by generating sets of realizations of the best-fit curve $C(t)$ \cite{Efron1}. Each realization of the cumulative number of case notifications $C_i(t)$ ($i=1$, $2$, $\ldots$, $m$) is generated as follows: for each observation $C(t)$ for $t=2$, $3$, $\ldots$, $n$ days generate a new observation $C_i^{'}(t)$ for $t \ge 2$ ($C^{'}_i(1)=C(1)$) that is sampled from a \textit{Poisson} distribution with mean: $C(t)-C(t-1)$ (the daily increment in $C(t)$ from day $t-1$ to
day $t$). The corresponding realization of the cumulative number of influenza notifications is given by $C_i(t) = \sum_{j=1}^{t} C^{'}_i(t)$ where $t=1$, $2$, $3$, $\ldots$, $n$. The reproduction number was then estimated from each of $1000$ simulated epidemic curves to generate a distribution of $R$ estimates from which simple statistics can be computed including $95\%$ confidence intervals. These statistics need to be interpreted with caution. For example, $95\%$ confidence intervals for $R$ derived from our bootstrap sample of $R$ should be interpreted as containing $95\%$ of future estimates when the same assumptions are made and the only noise source is observation error. It is tempting but incorrect to interpret these confidence intervals as containing the {\it true} parameters with probability $0.95$. Figure \ref{figR-SEIJR} shows the temporal distributions of the reproduction number and the proportion of the clinical reporting obtained by the bootstrap method after fitting the complex SEIR epidemic model to the initial phase of the Fall influenza wave using 17 epidemic days of the Spanish Flu Pandemic in San Francisco, California.

\section{Primer of mathematics and statistics of $R_0$ and $R(t)$}
In addition to the estimation of $R_0$, it is of practical importance to evaluate time-dependent variations in the transmission potential. Explanation of the time course of an epidemic can be partly achieved by estimating the effective reproduction number, $R(t)$, defined as the actual average number of secondary cases per primary case at time $t$ (for $t>0$) \cite{Wallinga1}\cite{Cauchemez1}\cite{Nishiura2006}\cite{Haydon004}\cite{Cauchemez2006EID}\footnote{$R(t)$ should not be confused with the number of removed individuals using the same notation. In the following arguments of this paper, $R(t)$ denotes the effective reproduction number.}. Although effective interventions against Spanish influenza may have been limited in the early 20th century, it is plausible that the contact frequency leading to infection varied with time owing to the huge number of deaths and dissemination of information through local media ({\it e.g.} newspapers). $R(t)$ shows time-dependent variation with a decline in susceptible individuals (intrinsic factors) and with the implementation of control measures (extrinsic factors). If $R(t)<1$, it suggests that the epidemic is in decline and may be regarded as being {\it under control} at time $t$ (vice versa, if $R(t)>1$). 

\subsection{Estimation of $R_0$ using the intrinsic growth rate}
\begin{figure}
   \begin{center}
   {\includegraphics[width=12cm, height=5.531cm]{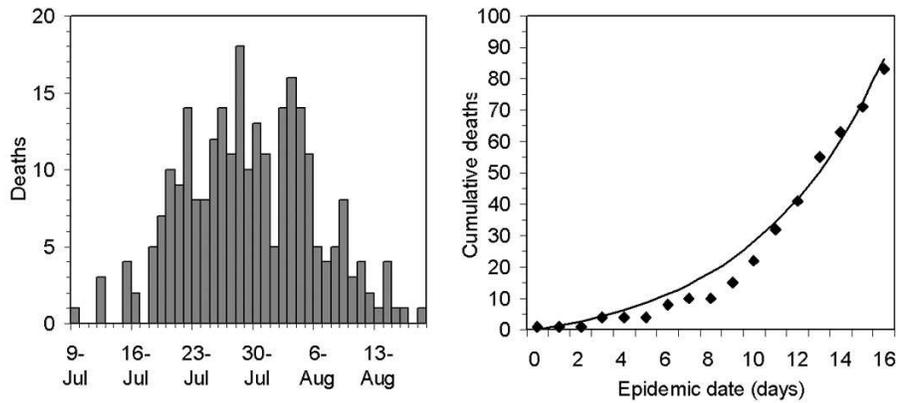}}
   \end{center}
   \caption{ {\bf Temporal distribution of Spanish influenza in Zurich.} Left panel shows an epidemic curve ({\it i.e.} deaths distribution) of pandemic influenza in a suburb of Zurich in 1918. In total, 259 deaths were observed from 9 July to 18 August. Right panel shows observed and expected values of the cumulative number of deaths during the first 16 days. The intrinsic growth rate is estimated to be 0.16 per day. Data source: \cite{Imahorn1919}}
\label{HN_fig1}
\end {figure}

To appropriately understand the theoretical concept of $R(t)$, let us firstly consider an explicitly infection-age structured epidemic model. Whereas Kermack-McKendrick model governed by ODEs (i.e. SIR and SEIR models as discussed above) has been well-known, Kermack and McKendrick had actually proposed an infection-age structured model in their initial publication in 1927 \cite{Kermack1927}, the mathematical importance of which was recognized only after 1970s \cite{Diekmann1977}\cite{Metz1978}. Let us denote the numbers of susceptible and recovered individuals by $S(t)$ and $U(t)$. Further, let $i(t, \tau)$ be the density of infectious individuals at time $t$ and {\bf infection-age} $\tau$  ({\it i.e.} time since infection). The model is given by 
\begin{equation}
\label{eqn_HN1}
\begin{array}{lcl}
\dfrac{dS(t)}{dt} &=& -\lambda(t) S(t)\\
\left(\dfrac{\partial}{\partial t}+\dfrac{\partial}{\partial \tau}\right)i(t,\tau) &=& -\gamma (\tau)i(t,\tau)\\
i(t,0)&=& \lambda(t) S(t)\\
\dfrac{dU(t)}{dt} &=& \int_{0}^{\infty} \gamma(\tau)i(t,\tau)\, d\tau
\end{array}
\end{equation}
where $\lambda(t)$ is referred to as the force of infection (foi) ({\it i.e.} as discussed in Section 2, foi is defined as the rate at which susceptible individuals get infected) which is given by:
\begin{equation}
\label{eqn_HN2}
\lambda(t) = \int_{0}^{\infty} \beta(\tau)i(t,\tau)\,d\tau
\end{equation}
and $\gamma(\tau)$ is the rate of recovery at infection-age $\tau$. It should be noted that the above model has not taken into account the background host demography ({\it i.e.} birth and death). In a closed population, the total population size $N$ is thus given by:
\begin{equation}
\label{eqn_HN3}
N = S(t)+\int_{0}^{\infty} i(t,\tau)\,d\tau + U(t)
\end{equation}
It should also be noted that, although $i(t,\tau)$ is referred to as {\it density}, it is not meant to be a normalized density ({\it i.e.} integral of $i(t,\tau)$ over $t$ and $\tau$  does not sum up to 1). Rather, we use density to mathematically refer to the absolute frequency in the infection-age space.
\\
\indent The system (\ref{eqn_HN1}) can be reasonably integrated
\begin{equation}
\label{eqn_HN4}
i(t,\tau) = 
\begin{cases} 
  \Gamma(\tau)j(t-\tau),  & for \quad t-\tau>0 \\
  \dfrac{\Gamma(\tau)}{\Gamma(\tau-t)}j_0(\tau-t), & for \quad \tau-t>0 
\end{cases}
\end{equation}
where
\begin{equation}
\label{eqn_HN5}
\begin{array}{lcl}
j(t) &=& i(t,0)\\
\Gamma(\tau) &=& \exp \left( - \int_{0}^{\tau} \gamma(\sigma)\,d\sigma \right)
\end{array}
\end{equation}
and $j_0(\tau)$ suggests the density of initially infected individuals at the beginning of an epidemic. In the following arguments, we call $j(t)$ as {\bf incidence of infection} ({\it i.e.} new infections at a given point of time $t$). It is not difficult to derive
\begin{equation}
\label{eqn_HN6}
S(t) = S(0)-\int_{0}^{t} j(\sigma)\,d\sigma
\end{equation}
from (\ref{eqn_HN1})- (\ref{eqn_HN5}). Thus, the subequation of $i(t,0)$ in system (\ref{eqn_HN1}) is rewritten as
\begin{equation}
\label{eqn_HN7}
j(t) = \lambda(t) \left[ S(0)-\int_{0}^{t} j(\sigma)\,d\sigma\right] \quad 
\end{equation}
Taking into account the initial condition in (\ref{eqn_HN4}), equation (\ref{eqn_HN6}) is rewritten as
\begin{equation}
\label{eqn_HN8}
j(t) = \left[ S(0)-\int_{0}^{t} j(\sigma)\,d\sigma\right]\left[ G(t)+\int_{0}^{t} \psi(\tau)j(t-\tau)\,d\tau \right] \quad 
\end{equation}
where
\begin{equation}
\label{eqn_HN9}
\begin{array}{lcl}
\psi(\tau) &=& \beta(\tau)\Gamma(\tau)\\
G(t) &=& \int_{0}^{\infty} \beta(\sigma+t)\dfrac{\Gamma(\sigma+t)}{\Gamma(\sigma)}j_0(\sigma)\,d\sigma 
\end{array}
\end{equation}
Considering the initial invasion phase ({\it i.e.} initial growth phase of an epidemic), we get a linearized equation
\begin{equation}
\label{eqn_HN10}
j(t) = S(0)G(t)+S(0) \int_{0}^{t}\psi(\tau)j(t-\tau)\,d\tau
\end{equation}
The equation (\ref{eqn_HN10}) represents Lotka's integral equation, where the basic reproduction number, $R_0$, is given by
\begin{equation}
\label{eqn_HN11}
R_0 = S(0) \int_{0}^{\infty}\psi(\tau)\,d\tau
\end{equation}
Thus, the epidemic will grow if $R_0>1$ and decline to extinction if $R_0<1$. The above model can yield the same final size equation as seen in models governed by ODEs \cite{Diekmann2000book}. 
\\
\indent Assuming that the infection-age distribution is stable, we get a simplified renewal equation
\begin{equation}
\label{eqn_HN12}
j(t) = \int_{0}^{\infty}A(\tau)j(t-\tau)\,d\tau
\end{equation}
where $A(\tau)$ is the product of $\psi(\tau)$ and $S(0)$. Moreover, assuming that we observe an exponential growth of incidence during the initial phase ({\i.e.} $j(t)=k\exp(rt)$ where $k$ and $r$ are, respectively, a constant ($k>0$) and the intrinsic growth rate), the following relationship must be met
\begin{equation}
\label{eqn_HN13}
j(t) = j(t-\tau)\exp(r\tau)
\end{equation}
Replacing $j(t-\tau)$ in the right hand side of (\ref{eqn_HN12}) by (\ref{eqn_HN13}), we get
\begin{equation}
\label{eqn_HN14}
j(t) = \int_{0}^{\infty}A(\tau)j(t)\exp(-r\tau)\,d\tau
\end{equation}
Removing $j(t)$ from both sides of (\ref{eqn_HN14}), we get the Lotka-Euler characteristic equation:\begin{equation}
\label{eqn_HN15}
1=\int_{0}^{\infty}e^{-r\tau}A(\tau),d\tau
\end{equation}
Further, we consider a probability density of the {\bf generation time} ({\it i.e. }the time from infection of an individual to the infection of a secondary case by that individual \cite{Svensson2007}), denoted by $w(\tau)$:
\begin{equation}
\label{eqn_HN16}
w(\tau):=\frac{A(\tau)}{\int_{0}^{\infty}A(x)dx}=\frac{A(\tau)}{R_0}.
\end{equation}
Using (\ref{eqn_HN16}), the equation (\ref{eqn_HN15}) can be replaced by
\begin{equation}
\label{eqn_HN17}
\dfrac{1}{R_0}=\int_{0}^{\infty}\exp(-r\tau)w(\tau),d\tau
\end{equation}
The equations (\ref{eqn_HN13})-(\ref{eqn_HN17}) are what Wallinga and Lipsitch discussed in a recent study \cite{Wallinga2}, reasonably suggesting the relationship between the generation time and $R_0$. Accordingly, the estimator of $R_0$ using the intrinsic growth rate is given by:
\begin{equation}
\label{eqn_HN18}
\hat{R_0} =\frac{1}{M(-r)},
\end{equation}
where $M(-r)$ is the moment generating function of the generation time distribution $w(\tau)$, given the intrinsic growth rate $r$ \cite{Wallinga2} \footnote{In the original study of Wallinga and Lipsitch \cite{Wallinga2}, the notation $R_0$ is not used for equation (\ref{eqn_HN18}) and rather document (\ref{eqn_HN18}) as the estimator of $R$. Most likely, there are two reasons for this. First, we cannot assure if all individuals are susceptible to pandemic influenza before the start of epidemic (as discussed in Section 2). Second, we cannot assume that infection-age distribution is stable during the initial growth phase, which is highlighted in (\ref{eqn_HN4}). Thus, it should be remembered that the above discussion is mathematically tight in theory, but there are certain number of assumptions to apply the concept to observed data. Since writing $R$ alone is always confusing (as it is unclear if $R$ is concerned with time or immunity status), here we use $R_0$ instead.}. Equation (\ref{eqn_HN18}) significantly improved the issue of estimating $R_0$ using the intrinsic growth rate alone, because (\ref{eqn_HN18}) permits validating estimates of $R_0$ by various different distributional assumptions for $w(\tau)$. The issue of assuming explicit distributions for latent and infectious periods has been highlighted in recent studies \cite{Wearing2005}\cite{Lloyd2001a}\cite{Lloyd2001b}\cite{Roberts2007},\cite{Yan1} and indeed, this point is in part addressed by (\ref{eqn_HN18}), because the convolution of latent and infectious periods yields $w(\tau)$. Moreover, since the assumed lengths of generation time most likely yielded different estimates of $R_0$ for Spanish influenza by different studies \cite{Nishiura1}, equation (\ref{eqn_HN18}) highlights a critical need to clarify the generation time distribution using observed data.
\\
\indent Here we briefly show a numerical example. Figure \ref{HN_fig1} shows the daily number of influenza deaths during Spanish influenza pandemic in a suburb of Zurich, 1918 \cite{Imahorn1919}. Since the non-linear phase is difficult to analyze, our interest to estimate $R_0$ with this dataset is limited to the initial growth phase only (right panel in Fig \ref{HN_fig1}). Even though the data represent deaths over time ({\it i.e.} not infection events with time), we can directly extract the same intrinsic growth rate as practised with onset data, assuming that death data are a good proxy for morbidity data (see our discussions in Section 6). Assuming exponential growth in deaths as shown in (\ref{eqn_HN13}), the intrinsic growth rate $r$ is estimated to be 0.16 per day. Supposing that $w(\tau)$ is arbitrarily assumed to follow a gamma distribution with mean $G$ and coefficient of variation, $CV=\sqrt{Var(G)}/G$, $R_0$ is given by
\begin{equation}
\label{eqn_HN189}
R_0 = (1+r G (CV)^2)^{\frac{1}{(CV)^2}}
\end{equation}
Although there is no concensus regarding the generation time of Spanish influenza, we assume it ranges from 2-5 days. Assuming further that $CV=0.5$, $R_0$ is estimated to range from 1.36 (for $G=2$ day) to 2.07 (for $G=5$ days).

\subsection{The concept of $R(t)$ and its estimation}
In the following, let us consider the non-linear phase of an epidemic. Derivation of $R_0$ given by (\ref{eqn_HN18}) assumes an exponential growth which is applicable only during the very initial phase of an epidemic (or, when the transmission is stationary over time), and thus, it is of practical importance to widen the utility of above-described renewal equations in order to appropriately interpret the time-course of an influenza pandemic. Let us explicitly account for the depletion of susceptible individuals, as we deal with an estimation issue with time-inhomogeneous assumptions ({\i.e.} non-linear phase). Adopting the {\bf mass action} assumption, we get:
\begin{equation}
\label{eqn_HN182}
\begin{aligned}
j(t) & = S(t)\int_{0}^{\infty} \psi(\tau)j(t-\tau)\,d\tau \\
      & = \int_{0}^{\infty}A(t,\tau)j(t-\tau)\,d\tau \\
\end{aligned}
\end{equation}
where $A(t,\tau)$ should be interpreted as the reproductive power at time $t$ and infection-age $\tau$ at which an infected individual generates secondary cases. We refer to the latter part of equation (\ref{eqn_HN182}) as a non-autonomous renewal equation, where the number of new infection at time $t$ is proportional to the number of infectious individuals (as assumed in the renewal equation in the initial phase). 
\\
\indent Using equation (\ref{eqn_HN182}), the effective reproduction number, $R(t)$ ({\it i.e.} instantaneous reproduction number at calendar time $t$) is defined as:
\begin{equation}
\label{eqn_HN19}
R(t) = \int_{0}^{\infty}A(t,\tau)\,d\tau
\end{equation}
Following (\ref{eqn_HN19}), we can immediately see that $R(t)$ with an autonomous assumption ({\it i.e.} where contact and recovery rates do not vary with time) is given by:
\begin{equation}
\label{eqn_HN20}
R(t) = \dfrac{S(t)}{S(0)}R_0
\end{equation}
which is shown in \cite{Diekmann2000book}. In practical terms, equation (\ref{eqn_HN20}) suggests that time-varying decrease in transmission potential as well as decline in the epidemic reflects only depletion of susceptible individuals. This corresponds to a classic assumption of the Kermack and McKendrick model.
\\
\indent However, as we discussed in the beginning of this section, we postulate that human contact behaviors (and other extrinsic factors) modifies the dynamics of pandemic influenza, assuming that the decline in incidence does reflect not only depletion of susceptibles but also various extrinsic dynamics ({\it e.g.} isolation, quarantine and closure of public buildings). Thus, instead of the assumption in (\ref{eqn_HN182}), we shall assume time-inhomogeneous $\psi(t,\tau)$; {\it i.e.}
\begin{equation}
\label{eqn_HN21}
\begin{aligned}
j(t) & = S(t)\int_{0}^{\infty} \psi(t,\tau)j(t-\tau)\,d\tau \\
      & = \int_{0}^{\infty}A(t,\tau)j(t-\tau)\,d\tau \\
\end{aligned}
\end{equation}
to describe $A(t,\tau)$. 
\\
\indent To derive simple estimator of $R(t)$, it is convenient to assume separation of variables for $A(t,\tau)$ (implicitly assuming that the relative infectiousness to infection-age is independent of calendar time) \cite{Fraser2007}. Under this assumption, $A(t,\tau)$ is rewritten as the product of two functions $\phi_1(t)$ and $\phi_2(\tau)$:
\begin{equation}
\label{eqn_HN22}
A(t,\tau) = \phi_1(t)\phi_2(\tau)
\end{equation}
Arbitrarily assuming a normalized density for $\phi_2(\tau)$, {\it i.e.},
\begin{equation}
\label{eqn_HN23}
\int_{0}^{\infty}\phi_2(\tau)\, d\tau \equiv 1
\end{equation}
then, it is easy to find that
\begin{equation}
\label{eqn_HN24}
R(t) = \int_{0}^{\infty}A(t,\tau)\,d\tau=\phi_1(t)
\end{equation}
suggesting that the function $\phi_1(t)$ is equivalent to the effective reproduction number $R(t)$. Another function $\phi_2(\tau)$ represents the density of infection events as a function of infection-age $\tau$. Accordingly, we can immediately see that $\phi_2(\tau)$ is exactly the same as $w(\tau)$, the generation time distribution. That is, the above arguments suggest that $A(t,\tau)$ ({\it i.e.} the rate at which an infectious individual at calendar time $t$ and infection-age $\tau$ produces secondary transmission) is decomposed as:
\begin{equation}
\label{eqn_HN25}
A(t,\tau) = R(t)w(\tau)
\end{equation}
Inserting (\ref{eqn_HN25}) into (\ref{eqn_HN21}) yields an estimator of $R(t)$ \cite{Fraser2007}:
\begin{equation}
\label{eqn_HN26}
\hat{R}(t) = \dfrac{j(t)}{\int_{0}^{\infty}j(t-\tau)w(\tau)\,d\tau}
\end{equation}
The above equation (\ref{eqn_HN26}) is exactly what was proposed in applications to SARS \cite{Wallinga1} and foot and mouth disease \cite{Ferguson2001}; {\it i.e.} discretizing (\ref{eqn_HN26}) to apply it to the daily incidence data ({\it i.e.} using $j_i$ incident cases infected between time $t_i$ and time $t_{i+1}$ and descretized generation time distribution $w_i$),
\begin{equation}
\label{eqn_HN27}
\hat{R}(t_i) = \dfrac{j_i}{\sum_{j=0}^{n}j_{i-j}w_j}
\end{equation}
was used as the estimator. However, it should be noted that the study in SARS implicitly assumed that onset data $c(t)$ at time $t$ reflects the above discussed infection event $j(t)$. That is, supposing that we observed $c_i$ onset cases reported between $t_i$ and $t_{i+1}$, $R(t)$ was calculated as
\begin{equation}
\label{eqn_HN28}
\hat{R}(t_i) = \dfrac{c_i}{\sum_{j=0}^{n}c_{i-j}s_j}
\end{equation}
where $s_j$ is the discretized {\bf serial interval} which is defined as the time from onset of a primary case to onset of the secondary cases \cite{Fine2003}\cite{Nishiura2006ije}. The method permits reasonable transformation of an epidemic curve ({\it i.e.} temporal distribution of case onset) to the estimates of time-inhomogeneous reproduction number $R(t)$. Employing the relative likelihood of case $k$ infected by case $l$ using the density function of serial interval $s(t)$; {\it i.e.},
\begin{equation}
\label{eqn_HN29}
p_{(k,l)} = \dfrac{s(t_k-t_l\vert\theta)}{\sum_{m\neq k}^{}s(t_k-t_m\vert\theta)}
\end{equation}
Using (\ref{eqn_HN29}), expected value and variance of $R(t_i)$ are given by the following
\begin{equation}
\label{eqn_HN30}
\begin{aligned}
E(R(t_i)) & = \dfrac{1}{n_t^2} \sum_{l:t_l=t}^{}\sum_{k=1}^{n-q}p_{(k,l)} \\
Var(R(t_i)) & = \dfrac{1}{n_t^2} \sum_{k=1}^{n-q} \left( \sum_{l:t_l=t}^{}p_{(k,l)}(1-p_{(k,l)})-\sum_{l,m:t_l=t_m=t}^{}p_{(k,l)}p_{(k,m)} \right)\\
\end{aligned}
\end{equation}
where $n_t$ is the total number of reported case onsets at time $t$ \cite{Cowling2007}. 
\\
\indent In the present day, only by using the above described methods (or similar concepts with similar assumptions), we can transform epidemic curves into $R(t)$ and roughly assess the impact of control measures on an epidemic. However, whereas the equations (\ref{eqn_HN27}) and (\ref{eqn_HN28}) are similar in theory, we need to explicitly account for the difference between onset and infection event. In fact, when there are many asymptomatic infections and asymptomatic secondary transmissions, serial interval is not equivalent to the generation time, and thus, directly adopting the above methods would be inappropriate. Since this point is particularly important in analyzing influenza data, we discuss this issue in Section 6.
\section{Statistical methods to estimate $R_0$}
\subsection{Branching process}
\indent In the previous sections, we discussed several different methods to estimate $R_0$ either by (i) employing detailed curve fitting method assuming a structured epidemic model or (ii) using the intrinsic growth rate (or doubling time \cite{Marques1994}\cite{Galvani2003}). Summarizing the above discussions, we believe that the readers should benefit from memorizing $R_0=1/M(-r)$ for the use of the intrinsic growth rate $r$ in estimating $R_0$ \cite{Wallinga2} and remembering the final size equation $R_0=-\ln (1-p)/p$ suggesting the severity of an epidemic as the theoretical concept. Indeed, estimator using either the intrinsic growth rate or final size has still continued to play an important role in discussing $R_0$ of pandemic influenza \cite{Vynnycky2007}. 
\\
\indent However, it should be noted that deterministic models do not permit incorporating stochasticity explicitly ({\it e.g.} standard error for $R_0$ is determined by measurement of errors alone), as the models argue only {\it average number of secondary transmissions} within the assumed transmission dynamics. That is, our arguments given above explore only the time-evolution of influenza spread in the {\bf mean field}. To address the issue of variation in secondary transmissions, full stochastic models are called for \cite{Lessler2007}. 
\\
\indent From a viewpoint of data science, the discrete-time branching process, which is also referred to as Galton-Watson process, can reasonably assess individual heterogeneity in secondary transmissions  \cite{Haccou2005}\cite{Kimmel2002}. As we discussed the initial growth phase of the deterministic model, let us consider the same epidemic phase where we observe a geometric increase in the number of cases by generation \cite{Nishiura2006Lotz}. We denote the initial number of infected individuals by $C_0$ in generation 0. Then, during the first generation, $C_1$ cases are produced by secondary transmissions of $C_0$. Similarly, let $C_n$ be the number of infections in generation $n$. The branching process of this type assumes that every infected individual has an independently and identically distributed stochastic random variable $\rho_i^{(n)}$ representing the number of secondary cases produced by case $i$ in generation ($n$), and that environmental stochasticity and immigration/emigration can be ignored. Supposing that the pattern of secondary transmission follows a discrete probability distribution $p_k$ with $k$ secondary transmission(s); {\it i.e.},
\begin{equation}
\label{eqn_HN202}
p_k = Pr(\rho_i^{(n)}=k) \quad  (k=0,1,2,...)
\end{equation}
then, the expected number of secondary transmissions and the variance are given by
\begin{equation}
\label{eqn_HN203}
\begin{array}{lcl}
R_0 &=& E(\rho_i^{(n)})=\sum_{k=1}^{\infty} kp_k\\
Var(\rho_i^{(n)}) &=&  \sum_{k=0}^\infty \left(k-R_0\right)^2p_k
\end{array}
\end{equation}
\indent In other words, the concept of probability distribution $p_k$ reflects {\bf offspring distribution} in population ecology, and this permits explicit modeling of variations in secondary transmissions in infectious diseases \cite{Becker1977a}\cite{Becker1977b}. This approach is particularly important during the initial phase of an epidemic, because the number of infectious individuals is small in this stage, and thus, it is deemed essential to take into account demographic stochasticity, {\it i.e.}, variation in the numbers of secondary transmissions by chance. Indeed, the model has been applied to observed outbreak data where we observed the extinction before growing to a major epidemic \cite{Farrington2003}\cite{Ferguson2004}.
\\
\indent Let us briefly discuss the variation in secondary transmissions and an estimation method of $R_0$ using the discrete-time branching process, deriving analytical expressions of the expected number of infected individuals in generation $n$, $M_n=E(C_n)$ and the variance $V_n=Var(C_n)$. It is impossible to avoid using the probability generating function (pgf) to discuss the branching process. The above described $\rho_i^{(n)}$ characterize {\it positive} and {\it discrete} number of secondary transmissions, and thus, is a non-zero discrete random variable. The pgf of $\rho$, $g_\rho(s)$ is given by
\begin{equation}
\label{eqn_HN205}
g_{\rho}(s) = E(s^{\rho})=\sum_{k=0}^{\infty} p_k s^k
\end{equation}
There are two basic properties concerning $g(s)$ in relation to the epidemic process. First, $R_0$ is by definition the mean value of secondary transmissions (equation (\ref{eqn_HN203})) and, thus given by $g'(1)$. Second, the probability that an infected individual does not cause any secondary transmissions $p_0=$ Pr($\rho$=0) is given by $g(0)$, which is useful for discussing threshold phenomena and extinction \cite{Kimmel2002}. If we note that $C_0=1$ (i.e. only one index case), the Galton-Watson process has the following pgf identity:
\begin{equation}
\label{eqn_HN206}
\begin{array}{lcl}
g_0 (s) &=& s\\
g_{n+1}(s) &=& g_n (g(s)) = g (g_n (s))
\end{array}
\end{equation}
Even when there are $C_0=a$ cases in generation 0 (where $a>1$), we just have to assume that there are $a$ different independent infection-trees and thus
\begin{equation}
\label{eqn_HN207}
\begin{array}{lcl}
g_{C_0} (s) &=& s^a\\
g_{C_{n}}(s) &=& (g_n (s))^a
\end{array}
\end{equation}
From the above discussions, the expected number of cases in generation $n$, $M_n$, and the variance $V_n$ is
\begin{equation}
\label{eqn_HN208}
\begin{array}{lcl}
M_n &=& R_0^n M_0\\
V_n &=& 
  \begin{cases} 
  n Var(\rho),  & (R_0=1) \\
  R_0^{n-1} Var(\rho) \dfrac{R_0^n-1}{R_0-1}, & (R_0 \ne 1) 
  \end{cases}
\end{array}
\end{equation}
The process grows geometrically if $R_0>1$, stays constant if $R_0=1$, and decays geometrically if $R_0<1$. These three cases are referred to as {\bf supercritical}, {\bf critical}, and {\bf subcritical}, respectively. However, unlike the deterministic model, it should be remembered that critical process does not permit continued transmissions, and rather, the process becomes extinct almost surely (i.e. probability of extinction given $R_0=1$ is one) \cite{Kimmel2002}. 
\\
\indent Mathematically, demographic stochasticity in transmission is represented by a Poisson process, which has been practiced in the application of branching processes to epidemics \cite{Diekmann2000book}. Assuming that mean value of secondary transmissions is a constant $R_0$, the conditional distribution of observing $C_{n+1}$ cases, given $C_{n}$ cases, follows a Poisson distribution:
\begin{equation}
\label{eqn_HN209}
C_{n+1} \vert C_n \sim  Poisson\left[ C_n R_0 \right]
\end{equation}
Supposing that we analyze influenza data documenting the generations of cases from 0 to $n$ in which we observed geometric growth, the likelihood of estimating $R_0$ is proportional to
\begin{equation}
\label{eqn_HN210}
\prod_{k=1}^n \left(R_0 C_{k-1} \right)^{C_k} \exp \left(-R_0 C_{k-1} \right)
\end{equation}
\indent Here we apply the above model to the Spanish influenza data in Zurich (Figure \ref{HN_fig1}). Assuming that the generation time of length $\tau$, $w(\tau)$, is given by the following delta function with the mean length 3 days,
\begin{equation}
\label{eqn_HN211}
w(\tau) = 
\begin{cases}
  \infty,  & for \quad \tau=3 \\
  0, & for \quad \tau\neq 3 
\end{cases}
\end{equation}
then the observed series of data can be grouped by generation ($C_0$, $C_1$, $C_2$, ...):
\begin{equation}
\label{eqn_HN212}
1, \quad 3, \quad 4, \quad 7, \quad 26, \quad 30, \quad 37, ...
\end{equation}
Since we assumed exponential growth during the initial 16 days in the previous section, here we similarly assume a geometric increase up to the 6th generation. Applying (\ref{eqn_HN210}) to the above data, maximum likelihood estimate of $R_0$ (and the corresponding 95 percent confidence intervals) is 1.51 (1.24, 1.81). The model is simple enough to estimate $R_0$, and indeed, a slight extension of the discrete-time branching process has been employed to estimate $R_0$ as well as the proportion of undiagnosed cases in the analysis of SARS outbreak data \cite{Glass2007}. 
\\
\indent It should be noted that the discrete-time branching process assumes homogeneous pattern of spread. A technical issue has arisen on this subject during the SARS outbreak. Usually, we observe some cases who produce an extraordinary number of secondary cases compared with other infected individuals, which are referred to as {\bf superspreaders}. Because of this, observed offspring distributions for directly transmitted diseases tend to be extremely skewed to the right. Empirically, it has been suggested that Poisson offspring distribution is sometimes insufficient to highlight the presence of superspreaders in epidemic modeling \cite{Lloyd-Smith2005}. For example, if non-zero discrete distribution of secondary cases follows a geometric distribution with mean $R_0$, the pgf is given by a geometric distribution with mean $R_0$:
\begin{equation}
\label{eqn_HN213}
g(s)=\dfrac{1}{1+R_0(1-s)}
\end{equation}
Moreover, if the offspring distribution follows gamma distribution with mean $R_0$ and dispersion parameter $k$, the pgf $g(s)$ follows negative binomial distribution \cite{Lloyd-Smith2007}:
\begin{equation}
\label{eqn_HN214}
g(s)=\left( 1 + \dfrac{R}{k}(1-s) \right)^{-k}
\end{equation}
We still do not know if pandemic influenza is also the case to warrant the skewed offspring distributions. To explicitly test if superspreading events frequently exist in influenza transmission, it is necessary to accumulate contact tracing data for this difficult disease, the cases of which often show flu-like symptoms only (as discussed in Section 1). In addition, it should be noted that we cannot attribute the skewed offspring distribution to the underlying contact network only. To date, there are two major reasons which can generate superspreaders: (i) those who experience very frequent contacts (social superspreader) or (ii) those who are suffering from high pathogen loads or those who can scatter the pathogen through the air such as the use of nebuliser in hospitals (biological superspreader). From the offspring distribution only, we cannot distinguish these two mechanisms.

\subsection{Counting process}
With regard to the estimation of $R_0$ using final size, we briefly discuss another method based on a stochastic process. As we discussed above, let $S(t)$, $I(t)$ and $U(t)$ be the numbers of susceptible, infectious and recovered individuals at time $t$, respectively. Further, let $\beta$ and $1/\gamma$ denote the transmission rate and the mean duration of the infectious period, respectively. Supposing that $K(t)$, the number of individuals who experienced infection between time 0 and time $t$, is given by $K(t)=S(0)-S(t)$, the two processes $K(t)$ and $U(t)$ are increasing counting processes where the general epidemic is explained by:
\begin{equation}
\label{eqn_HN215}
\begin{array}{lcl}
Pr(dK(t)=1,dU(t)=0\vert\mathbb{Z}_t) &=& \beta\bar{S}(t)I(t)dt\\
Pr(dK(t)=0,dU(t)=1\vert\mathbb{Z}_t) &=& \gamma I(t)dt\\
Pr(dK(t)=0,dU(t)=0\vert\mathbb{Z}_t) &=& 1-\beta\bar{S}(t)I(t)dt-\gamma I(t)dt
\end{array}
\end{equation}
where $\mathbb{Z}_t$ denotes the $\sigma$-algebra generated by the history of the epidemic $\left\lbrace S(u),I(u);0<u<t\right\rbrace$ and $\bar{S}(t)=S(t)/n$ (where $n$ is the size of the susceptible population at time 0). The latter is equivalent to assuming density-independent transmission ({\it i.e.} also referred to as {\it true mass action} or frequency dependent assumption \cite{DeJong1995}). Based on equation (\ref{eqn_HN215}), two zero-mean martingales \cite{Becker1993} are defined by:
\begin{equation}
\label{eqn_HN216}
\begin{array}{lcl}
M_1(t) &=& K(t)-\int_{0}^{t} \beta\bar{S}(u)I(u)\, du\\
M_2(t) &=& U(t)-\int_{0}^{t} \gamma I(u)\, du
\end{array}
\end{equation}
From the martingale theory \cite{Hall1980}, a zero-mean martingale is given by
\begin{equation}
\label{eqn_HN217}
\begin{array}{lcl}
M(t) &=& \int_{0}^{t}\dfrac{1}{\bar{S}(u)}\,dM_1(u) -\dfrac{\beta}{\gamma}M_2(t)  \\
 &=& \dfrac{n}{S(0)} + \dfrac{n}{S(0)-1}+ ...+ \dfrac{n}{S(t)+1}-\dfrac{\beta}{\gamma}U(t)
\end{array}
\end{equation}
Thus, the estimator $\hat{\theta}=\hat{\beta}/\hat{\gamma}$ is given by
\begin{equation}
\label{eqn_HN218}
\begin{array}{lcl}
\hat{\theta} &=& \dfrac{\left[\dfrac{n}{S(0)} + \dfrac{n}{S(0)-1}+ ...+ \dfrac{n}{S(T)+1}\right]}{U(T)}  \\
 &=& \dfrac{-ln(1-\tilde{p})}{U(T)}
\end{array}
\end{equation}
where $\tilde{p}$ is the observed final size ($=1-S(T)/n$) at the end of the epidemic at time $T$. Furthermore, the variance of the zero-martingale is given by
\begin{equation}
\label{eqn_HN219}
Var(M(t))=Var(M_1(t))+\theta^2Var(M_2(t))
\end{equation}
From the martingale central limit theorem \cite{Rida1991}, the estimator $\theta$ is approximately normally distributed in a major outbreak in a large community. The standard error is then consistently estimated by:
\begin{equation}
\label{eqn_HN220}
\begin{array}{lcl}
s.e.(\hat{\theta}) &=& \dfrac{\left[\dfrac{n}{S(0)^2} + \dfrac{n}{(S(0)-1)^2}+ ...+ \dfrac{n}{(S(T)+1)^2}-\hat{\theta}^2R(T) \right]^{\dfrac{1}{2}}}{U(T)}  \\
 &=& \dfrac{\left[\dfrac{n}{S(0)+\dfrac{1}{2}} + \dfrac{n}{S(0)+\dfrac{1}{2}}-\hat{\theta}^2R(T) \right]^{\dfrac{1}{2}}}{U(T)}
\end{array}
\end{equation}
Consequently, the estimator and standard error of $R_0$ are given by:
\begin{equation}
\label{eqn_HN221}
\begin{array}{lcl}
\hat{R_0} &=& n\times \hat{\theta}  \\
s.e.(\hat{R_0}) &=& n \times s.e.(\hat{\theta})
\end{array}
\end{equation}
More detailed mathematical descriptions can be found elsewhere \cite{Becker1989}\cite{Becker1999}\cite{Andersson2000}.
\\
\indent Here we show a numerical example. Let us consider a large epidemic of equine influenza ({\it i.e.} influenza in horses) as our case study. In 1971, a nationwide epidemic of equine-2 influenza A (H3N8) was observed in Japan \cite{Satou2006}. For example, in Niigata Racecourse, 580 influenza cases were diagnosed with influenza among a total of 640 susceptible horses. The final size $p$ is thus 90.6 percent (95 percent CI: 88.4, 92.9). From this data, we calculate $R_0$ and its uncertainty bounds.
\\
\indent Using $\tilde{p}=0.906$ and total number of infected $U(T)=580$ in equation (\ref{eqn_HN218}), $\hat{\theta}$ is estimated as 0.00408. Therefore, the estimate of $R_0= 2.60$ is given by equation (\ref{eqn_HN221}). Moreover, from equation (\ref{eqn_HN220}) where $S(T)=639-579=60$ and $S(0)(=n)=639$ (we assume one case was already infected at time $t=0$), we obtain $s.e.(\hat{\theta})=0.000126$. Here, $\hat{\theta}$ is assumed to follow normal distribution. Therefore, the 95 percent confidence interval for $R_0$ is given as $[R_0-1.96\times640\times0.000126, R_0+1.96\times640\times0.000126]=[2.44,2.76]$. 
\\
\indent When applying the final size equation, it should be remembered that (i) we assume all individuals are initially susceptible (in the above described model) and (ii) we assume $\beta$ and $\gamma$ are independent of time ({\it i.e.} constant), and thus, that any extrinsic factors should not have influenced the course of the observed epidemic.

\subsection{Epidemics with two levels of mixing}
In the above described models, we always assumed that the pattern of influenza transmission is homogeneous, which is clearly unrealistic to capture the transmission dynamics of influenza. Since the last century, it has already been understood that the transmission dynamics are not sufficiently modeled by assuming homogeneous mixing. However, because more detailed data are lacking ({\it e.g.} epidemic records of pandemic influenza with time, age and space), what we could offer has been mainly to extract the intrinsic growth rate from the initial exponential growth, and estimate $R_0$ using the estimator based on a model with the homogeneous mixing assumption. 
\\
\indent One line of addressing heterogeneous patterns of transmission using the observed data is separating household transmission from community transmission. In other words, it is of practical importance to distinguish between individual and group $R_0$ \cite{Barnes2007}. From the beginning of explicit modeling of influenza \cite{Longini1982a}\cite{Longini1982b}, a method to separately estimate the transmission parameters has been proposed, which has been partly extended in a recent study \cite{Fraser2007} or applied to further old data of pandemic influenza \cite{Nishiura2007flu}. Indeed, an important aspect of this issue was highlighted in a recent study which compared estimates of $R_0$ between those having casual and close contacts \cite{Vynnycky2007}. To estimate key parameters of household and community transmissions of influenza, or to simulate realistic patterns of influenza spread, such a consideration is fruitful.
\\
\indent Mathematically elaborating this concept, there are several publications which proposed the basis of analyzing household transmission data employing stochastic models \cite{Ball1997}\cite{Ball2002}\cite{Becker1995}. Moreover, a rigorous study has been made to estimate parameters determining the intrinsic dynamics ({\it e.g.} infectious period) using household transmission data with time \cite{Cauchemez2004}.
\\
\indent Future challenges on the estimation of $R_0$ include the application of such theories to the observed data with some extension. For example, as we discussed above, knowing the generation time would be crucial to elucidate a robust estimate of $R_0$ \cite{Wallinga2}\cite{Roberts2007}\cite{Fraser2007}\cite{Yan1}. However, we do not know if the generation time varies between close and casual contacts; this should be the case, because, as long as the generation time is given by covolution of latent and infectious periods, close contact should lead to shorter generation time than casual contact. In future studies, influenza models may better to highlight the increasing importance of considering household transmission to estimate the transmission potential using the temporal distribution of infection events. 

\section{Characteristics of influenza data}
Except for our approach in Section 3, mathematical arguments given in this paper are not particularly special for influenza. In other words, we modelers have employed similarly structured models which describe the population dynamics of other directly transmitted diseases, and such models are applicable not only for influenza but also for many viral diseases including measles, smallpox, chickenpox, rubella and so on \cite{Anderson1991}. However, influenza has many different epidemiologic characteristics compared to other childhood viral diseases. For instance, following the previous efforts in influenza epidemiology \cite{Kilbourne1987}\cite{Kilbourne1975}\cite{Hope-Simpson1992} and modeling \cite{Selby1982}\cite{Cliff1986}, we should at least note the following:

\begin{enumerate}
\item Detailed mechanisms of immunity have yet to be clarified. Since influenza virus has an wide antigenic diversity ({\it i.e.} unlike other childhood viral diseases, antigenic stimulation is not monoclonal), this complicates our understanding in the fraction of immune individuals, cross-protection mechanisms and evolutionary dynamics \cite{Ferguson2003}\cite{Abu-Raddad2004}\cite{Nelson2006}\cite{Koelle2006}.
\item Flu-like symptoms are too common, and thus, we cannnot explicitly distinguish influenza from other common viral infections without expensive laboratory tests for each case. Because of this character, it is difficult to effectively implement usual public health measures ({\it e.g.} contact tracing and isolation). 
\item Although explicit estimates are limited \cite{Vynnycky2007Asian}\cite{Mulder1958}, a certain fraction of infected individuals does not exhibit any symptoms (following infection). This complicates not only the eradication \cite{Fraser2004} but also epidemiologic evaluations of vaccines and therapeutics \cite{Kemper1980}.
\item Looking into the details of the intrinsic dynamics \cite{Ferguson1}\cite{Cauchemez2004}, it appears recently that the generation time and infectious period are much shorter than what were believed previously. Therefore, despite the relatively small $R_0$ estimate, the turn-over of a transmission cycle ({\it i.e.} speed of growth) is rather quick. The incubation period of Spanish influenza is as short as 1.5 days \cite{Nishiura2007incubation}, complicating the implementation of quarantine measures \cite{Day2006}.
\end{enumerate}
Thus, depending on the characteristics of observed data (and the specific purpose of modeling), we have to highlight these factors referring to the best available evidence. This is one of the most challenging issues in designing public health interventions against a potential future pandemic. 

\subsection{What is reported in the observed data?}
In addition to the above described issue, we, of course, must remember what the reported data is. In many studies, the compartment $I(t)$ or relevant class of infectious individuals of the SIR (or SEIR) model was fitted to the observed data. Indeed, in the majority of previous classic studies, $R(t)$ ({\it i.e.} removed class; denoted by $U(t)$ in our discussion) of Kermack and McKendrick model was fitted to the data, assuming that the removed class highlights observed data as the reported cases no longer produce secondary cases. However, the observed epidemiologic data is actually neither $I(t)$ nor $U(t)$. Always, what we get as the temporal distribution reflects {\it case onset} or {\it deaths} with time which is mostly accompanied by some reporting delay.
\\
\indent We believe this is one of the most challenging issues in epidemic modeling. Except for rare examples in sexually transmitted diseases, infection event is not directly observable, and thus, we have to maximize the utility of reported (observed) data, explicitly understanding what the data represents.
\\
\indent In this case, back-calculation of the infection events is called for. Let $c(t)$ denote the number of onsets at time $t$, this should be modeled by using incidence $j(t)$ and the density of the incubation period of length $\tau$, $f(\tau)$:
\begin{equation}
\label{eqn_HN301}
c(t)=\int_{0}^{\infty}  j(t-\tau)f(\tau)\, d\tau
\end{equation}
Further, supposing that $b(t)$ is the number of reported cases at time $t$ and the density of reporting delay of length $\sigma$ is $h(\sigma)$, observed data is modeled as:
\begin{equation}
\label{eqn_HN302}
\begin{array}{lcl}
b(t) &=& \int_{0}^{\infty}  c(t-s)h(s)\, ds  \\
     &=& \int_{0}^{\infty}  \int_{0}^{\infty}  j(t-s-\tau)f(\tau)\, d\tau h(s)\, ds
\end{array}
\end{equation}
That is, only by using the observed data $b(t)$ and known information of the reporting delay $h(s)$ and incubation period distributions $f(\tau)$, we can translate the observed data into infection process $j(t)$. 
\\
\indent In some cases, only death data with time, $d(t)$, is available \cite{Nishiura1}. Similarly, this can be modeled using the backcalculation. Let $q$ denote the case fatality of influenza which is reasonably assumed time-independent, and further let $m(u)$ be the relative frequency of time from onset to death, $d(t)$ is given by:
\begin{equation}
\label{eqn_HN303}
d(t)= q \int_{0}^{\infty}  c(t-u)m(u)\, du
\end{equation}
Even when using {\it onset data with delay} or {\it death} data, it should be noted that the intrinsic growth rate is identical to that estimated from the infection event distribution. Assuming that the incidence $j(t)$ exhibits exponential growth during the initial phase of an epidemic, {\it i.e.}, $j(t)=k\exp(rt)$, equations (\ref{eqn_HN301}) and (\ref{eqn_HN303}) can be rewritten as
\begin{equation}
\label{eqn_HN304}
\begin{array}{lcl}
b(t) &=& \int_{0}^{\infty}  \int_{0}^{\infty}  k\exp(r(t-s-\tau))f(\tau)\, d\tau h(s)\, ds  \\
     &=& k\exp(rt)\int_{0}^{\infty}  \int_{0}^{\infty}  \exp(-r(s+\tau))f(\tau)\, d\tau h(s)\, ds
\end{array}
\end{equation}
and
\begin{equation}
\label{eqn_HN305}
\begin{array}{lcl}
d(t) &=& q \int_{0}^{\infty}  \int_{0}^{\infty}  j(t-u-\tau)f(\tau)\, d\tau m(u)\, du \\
     &=& q \int_{0}^{\infty}  \int_{0}^{\infty}  k\exp(r(t-u-\tau))f(\tau)\, d\tau m(u)\, du\\
     &=& qk\exp(rt) \int_{0}^{\infty}  \int_{0}^{\infty}  \exp(-r(u+\tau))f(\tau)\, d\tau m(u)\, du
\end{array}
\end{equation}
Thus, the growth terms $\exp(rt)$ ({\it i.e.} which depends on time) of $b(t)$ and $d(t)$ are still identical to that of incidence $j(t)$. In other words, mathematically the equations (\ref{eqn_HN304}) and (\ref{eqn_HN305}) could be a justification to extract an estimate of the intrinsic growth rate from cases with reporting delay or deaths with time. However, we should always remember that the infection-age distribution is not stable during the initial phase, and moreover, this method cannot address individual variation in the secondary transmissions ({\it e.g.} superspreaders, as we discussed in Section 5). 

\subsection{What to be learnt from the reported data?}
In this way, it's not an easy task to clarify the infection events with time. A similar application of the convolution equation has been intensively studied in modeling HIV/AIDS. Since AIDS has a long incubation period, and because AIDS diagnosis is certainly reported in the surveillance data (at least, in industrialized countries), backcalculation of the number of HIV infections with time using the nubmer of AIDS diagnoses and the incubation period distribution has been an issue to capture the whole epidemiologic picture of HIV/AIDS \cite{Brookmeyer1994}\cite{Jewell1992}\cite{Colton1994}\cite{Nishiura2007HIV}. In the current modeling practice using the temporal distribution of onset events, we are now faced with a need to apply this technique to diseases with much shorter incubation periods.
\\
\indent Now, let's look back at a method to estimate $R(t)$, which was proposed by Wallinga and Teunis \cite{Wallinga1}. Whereas the method has a background of mathematical reasoning (as shown in (\ref{eqn_HN28}), Section 4.2), the estimator was derived implicitly assuming that {\it observed data exactly reflects infection events}. If asymptomatic infection and transmission are rare, this assumption might be justifiable as the lengths of the serial interval and generation time become almost identical. However, as long as we cannot ignore asymptomatic transmissions, which is particularly the case for influenza, the assumption $s(\sigma)=w(\tau)$ might be problematic \cite{Svensson2007}.
\\
\indent Since $R(t)$ of this method was given by summing up the probability of causing secondary transmissions by an onset case {\it at the onset time} of this case $t$, we should rewrite the assumption using a modified {\it onset-based} renewal equation as follows \cite{Nishiura1}:
\begin{equation}
\label{eqn_HN306}
c(t)=\int_{0}^{t}  c(t-\tau)R(t-\tau)s(\tau)\, d\tau
\end{equation}
For simplicity, we ignore reporting delay in the observed data, roughly assuming that the observed data reflects $c(t)$. Translating equation (\ref{eqn_HN306}) in words, it is implicitly assumed that {\it secondary transmission happens exactly at the time of onset}, and based on this assumption, $R(t)$ in the right hand side of (\ref{eqn_HN306}) can be backcalculated.
\\
\indent To understand the assumptions behind the above equation, let us assume that incidence $j(t)$ is given by
\begin{equation}
\label{eqn_HN307}
j(t)=S(t) \int_{0}^{\infty}  \beta(\sigma)\Gamma(\sigma)c(t-\sigma)\, d\sigma
\end{equation}
where $\beta(\sigma)$ is the transmission rate at {\bf disease-age} $\sigma$ ({\it i.e.} the time since onset of infection \cite{Nishiura2007smlx}) and $\Gamma(\sigma)$ is the survivorship of cases following onset. It should be noted that equation (\ref{eqn_HN307}) ignores secondary transmissions before onset of illness. As we discussed above, $c(t)$ is given by $j(t)$ and the incubation period distribution $f(\tau)$,
\begin{equation}
\label{eqn_HN308}
c(t)=\int_{0}^{\infty}  j(t-\tau)f(\tau)\, d\tau
\end{equation}
Replacing $c(t)$ in the right hand side of (\ref{eqn_HN307}) by (\ref{eqn_HN308}), we get
\begin{equation}
\label{eqn_HN309}
j(t)=S(t) \int_{0}^{\infty}  j(t-s)\phi(s)\, ds
\end{equation}
where $s$ represents infection-age ({\it i.e.} time since infection), and $\phi(s)$ is given by
\begin{equation}
\label{eqn_HN310}
\phi(s)=\int_{0}^{s} \beta(a)\Gamma(a)f(s-a)\, da
\end{equation}
which represents generation time distribution. From equations (\ref{eqn_HN309}) and (\ref{eqn_HN310}), we can find that $R(t)$ is given by
\begin{equation}
\label{eqn_HN311}
\begin{array}{lcl}
R(t) &=& S(t) \int_{0}^{\infty}  \phi(s)\, ds \\
     &=& S(t) \int_{0}^{\infty}  \beta(\sigma)\Gamma(\sigma)\,d\sigma \int_{0}^{\infty} f(\tau)\, d\tau 
\end{array}
\end{equation}
Equation (\ref{eqn_HN311}) can be further reduced to $R(t)=R_0 S(t)/S(0)$ which represents Kermack and McKendrick's assumption. Replacing $j(t)$ in the right hand side of (\ref{eqn_HN308}) by (\ref{eqn_HN307}), we get
\begin{equation}
\label{eqn_HN312}
c(t)=\int_{0}^{\infty} \psi(t,\sigma)c(t-\sigma)\, d\sigma
\end{equation}
where $\psi(t,\sigma)$ denotes the serial interval distribution of calender time $t$ and disease-age $\sigma$:
\begin{equation}
\label{eqn_HN313}
\psi(t,\sigma)=\int_{0}^{\sigma} \beta(\sigma-\tau)\Gamma(\sigma-\tau)f(\tau)S(t-\tau)\, d\tau
\end{equation}
Equation (\ref{eqn_HN313}) is difficult to solve as it includes $S(t-\tau)$ in the right hand side. However, in the special case, {\it e.g.}, let's say when we can assume $\beta(\tau)\Gamma(\tau)=k\delta(\tau)$ (where $k$ is constant and $\delta(\cdot)$ is delta function), 
\begin{equation}
\label{eqn_HN314}
\psi(t,\sigma)=kf(\sigma)S(t-\sigma)
\end{equation}
Inserting (\ref{eqn_HN314}) back to (\ref{eqn_HN312}),
\begin{equation}
\label{eqn_HN315}
\begin{array}{lcl}
c(t)&=&\int_{0}^{\infty} kf(\sigma)S(t-\sigma)c(t-\sigma)\, d\sigma \\
     &=& \int_{0}^{\infty}   R(t-\sigma)c(t-\sigma)\dfrac{f(\sigma)}{\int_{0}^{\infty} f(\tau)\,d\tau} \, d\sigma
\end{array}
\end{equation}
which is {\it onset-based} renewal equation which was presented in (\ref{eqn_HN306}). What to be learnt from (\ref{eqn_HN315}) is, the assumption that {\it secondary transmission happens immediately after onset} suggests that the {\it incubation period distribution is identical to the serial interval distribution} as shown above, which is a bit funny conclusion. Maximizing the utility of observed data has still remained an open question. \\
\indent In addition to modeling the temporal distribution, explicit modeling of asymptomatic infection is also called for \cite{Fraser2004}. Provided that there are so many asymptomatic transmissions which are not in the negligible order, we need to shift our concept of transmissibility; {\it e.g.}, rather than $R_0$, a threshold quantity of symptomatic infection is required. In such a case, application of type-reproduction number $T$ might be useful \cite{Heesterbeek2007}\cite{Roberts2003}, and it has already been put into practice \cite{Inaba2008}. 
\section{What to be clarified further?}
In this review, we focused on the use of the temporal distribution of influenza to estimate $R_0$ (or $R(t)$) and the relevant key parameters. It must be remembered that our arguments, almost necessarily, employed homogeneous mixing assumption, as we cannot extract information on heterogeneous patterns of infection from a single stream of temporal data alone. Presently, more information ({\it e.g.} at least, spatio-temporal distribution) is becoming available for influenza. In this section, we briefly sketch what can be (and should be) done in the future to quantify the transmission dynamics of pandemic influenza.

\subsection{What $R_0$ really means?}
It's not a new issue that heterogeneous patterns of transmission could even destroy the mean field theory in infectious diseases. For example, in a pioneering study of gonorrhea transmission dynamics by Hethcote and Yorke \cite{Hethcote1980}, an importance of contact heterogeneity was sufficiently highlighted. Since a simple model assuming homogeneous mixing did not reflect the patterns of gonorrhea transmission in the United States, Hethcote and Yorke divided the population in question into two; those who are sexually very active and not, the former of which was referred to as {\bf core} group. Compared with the temporal distribution of infection given by a model with homogeneous mixing assumption, the simple heterogeneous model with a core group revealed much quicker increase in epidemic size, showing rather different trajectory of an epidemic. Given that the variance of sexual partnership is extremely large ({\it i.e.} if the distribution of the frequency of sexual intercourse is extremely skewed to the right with a very long right tail), the estimate of $R_0$ is shown to become considerably high. The finding supports a vulnerability of our society to the invasion of sexually transmitted diseases. Following this pioneering study, considerable efforts have been made to approximately model the heterogeneous patterns of transmission using extended mean field equations \cite{Anderson1991}\cite{Diekmann2}\cite{Woolhouse1997}\cite{Blower1991}. 
\\
\indent In addition to such an approximation of heterogeneous transmission, recent progress in epidemic modeling with explicit contact network structures suggests that variance of the contact frequency plays a key role in determining the threshold quantity, and in some special cases, the concept of threshold phenomena could be confused \cite{Newman2002}\cite{May2001}\cite{Miller2007}\cite{Duerr2007}. In Section 4, we defined the force of infection as
\begin{equation}
\label{eqn_HN401}
\lambda(t) = \int_{0}^{\infty} \beta(\tau)i(t,\tau)\,d\tau
\end{equation}
In deteministic models given by simple ODEs (which ignores infection-age), $\lambda(t)$ is equivalent to $\beta I(t)$. These are what classical mean field models suggest. 
\\
\indent Let us account for an epidemic on networks, whose node-connectivity distribution ({\it i.e.} the distribution of probabilities that nodes have exactly $k$ neighbors) follows some explicit distribution $P(k)$. The force of infection $\lambda_c$, which yields $R_0=1$, in a static contact network is given by
\begin{equation}
\label{eqn_HN402}
\lambda_c = \dfrac{\langle k \rangle}{\langle k^2 \rangle}
\end{equation}
Here $\langle k \rangle$ denotes the average connectivity of the nodes. Assuming that $P(k)$ follows a power law of the form $P(k)=c k^{-v}$ (where $c$ is constant),
\begin{equation}
\label{eqn_HN403}
\langle k^2 \rangle = c\sum_{k}^{} k^{2-v}
\end{equation}
Given that $v\le3$, $\lambda_c=0$, and in such a case, $R_0$ even becomes infinite. This implies that the disease spread will continue for any mean estimate of $R_0$. Such a network structure is referred to as {\bf scale free} \cite{Albert2002}, complicating disease control efforts in public health \cite{Kiss2005}\cite{Kiss2006}. The importance of the network structure would also be highlighted for $v>3$. 
\\
\indent For sexually transmitted infections, contact frequency is countable (unlike airborne infection or transmission through droplets), and $v$ is estimated to be around 3 or a little larger \cite{Liljeros2001}. Following such a finding, many non-sexual directly transmitted diseases are also modeled in the present day assuming the scale-free network \cite{Newman2002}. However, it should be noted that the pattern of contact does not necessarily follow scale-free for all directly transmitted diseases. Indeed, there is no empirical evidence which suggests that the contact structure of any droplet infections follows the power law ({\it i.e.} we do not know if the above described contact heterogeneity is the case for diseases except for sexually transmitted diseases). A typical example of confusion is seen in the superspreading events during the 2002-03 SARS epidemics \cite{Masuda2004}, where we cannot explicitly attribute the phenomena either to contact network or biological factors (as long as {\it contact} and infection event are not directly observable). We still do not know how we should account for the distribution $P(k)$ for influenza and other viral respiratory diseases ({\it i.e.} power law or not) which remains to be clarified for each disease in future research. 

\subsection{New theory to replace mass action principle}
Methodoligical developments have been made to account for the network heterogeneity with data \cite{Keeling2005}. An approximate approach to address this issue is highlighted particularly in spatio-temporal modeling, an excellent account of which is reviewed by Matt Keeling \cite{Keeling2005book}. 
\\
\indent Even though it's difficult to quantify the transmission dynamics with an explicit contact network with time, there are useful analytical approximations to capture the dynamics of influenza (and other respiratory transmitted viral diseases) and estimate the transmission potential. For example, the force of infection with a power law approximation is reasonably given by:
\begin{equation}
\label{eqn_HN404}
\lambda(t) = \beta I(t)^{1+\alpha}S(t)^{1+\Psi}
\end{equation}
In (\ref{eqn_HN404}), $\alpha$ and $\Psi$ characterize the epidemic dynamics; {\it e.g.} initial growth ({\it i.e} if $\alpha$ is less than 0, the modified form acts to dampen the exponential growth of incidence) and endemic equilibrium ({\it i.e.} when $\Psi$ is greater than 0, density-dependent damping is increased). A model of this type was actually validated with measles data in England and Wales, comparing the prediction with that of employing the mass action principle \cite{Bjornstad2002}. 
\\
\indent Another approximation might be a pair-wise model \cite{Keeling1999}, which can explicitly account for the correlation between connected pairs. The model reasonably permits deriving the force of infection $\lambda$ using the number of various connected paris, which implies wide applicability to the epidemiologic data of sexually transmitted infections. Incorporating spatial heterogeneity in an approximate manner would shed light on further quantifications \cite{Riley2007}\cite{Keeling2004}, and thus, simple and reasonably tractable models which permit spatio-temporal modeling of influenza are expected ({\it e.g.} \cite{Viboud3}). 

\subsection{Which kind of data do we have to explore?}
Summarizing the above discussions, we have presented modeling approaches that can quantify the transmission potential of pandemic influenza. As we have shown, temporal case distributions have been analyzed in many instances, and previous efforts have come close to maximize the utility of temporal distributions ({\it i.e.} epidemic curve). However, at the same time, we have also learned that we can extract almost the intrinsic growth rate alone from a single time-evolution data. Accordingly, we are now faced with a need to clarify heterogeneous patterns of transmission and more detailed intrinsic dynamics of influenza \cite{NishiuraEID}\cite{Duerr2007b}\cite{Eichner2007}. With regard to the latter, primitive epidemiologic questions ({\it e.g.} probability of clinical attack given infection) remain to be answered for Spanish, Asian and Hong Kong influenza. Let's summarize what we need to clarify theoretically about pandemic influenza in list:

\begin{enumerate}
\item Acquired immunity
\item Evolutionary dynamics
\item Multi-host species transmission
\item Asymptomatic transmission
\item Attack rate ({\it i.e.} Pr(onset$\vert$infection))
\item Case fatality ({\it i.e.} Pr(death$\vert$onset))
\item Generation time and serial interval
\item Latent, incubation, infectious and symptomatic periods with further data
\item Transmission potential with time, space and antigenic types
\item Transmission potential with time and age
\end{enumerate}
These issues highlight an importance to quantify the transmission of influenza using not only cases ({\it i.e.} those followed onset of symptoms) but also some hint suggesting the infection event. For example, majority of the above listed issues could be reasonably addressed by implementing serological surveys ({\it e.g.} antibody titers of individuals and, preferably, time-delay delay distribution from infection to seroconversion). Since the proportion of those who do not experience symptomatic infection ({\it i.e.} probability of asymptomatic infection) is not small for influenza \cite{Mathews1}\cite{Inaba2008}, case records can tell us little to address the above mentioned issues, and thus, historical data of Spanish influenza may hardly offer further information. By maximizing the utility of observed data, we have to clarify the dynamics of influenza further, and identify key information which characterize the specific mechanisms of spread.

\begin{table}[hp*]
\caption{Reported estimates of $R_0$ for pandemic influenza during the fall wave (2nd wave) from 1918-19}

\begin {tabular}{l|c|c|c|c}
\hline
Location &  Serial interval & $R_0$ & Autonomous system & Reference  \\
 &  (days) & &  fitted with entire  &   \\
  &  & & epidemic curve  &  \\
\hline
 San Francisco, USA & 6 & 3.5 & Yes & \cite{Chowell2} \\
                                      & 6 & 2.4 & No & \cite{Chowell2} \\
 45 cities in the USA & 6$^a$ & 2.7 & No & \cite{Mills1} \\
 UK (entire England and Wales)$^b$ & 6 & 1.6 & Yes & \cite{Gani1} \\
 Scandinavian cities & 6 & 1.4-1.6 & No & \cite{Andreasen1} \\
 Geneva, Switzerland & 5.7 & 3.8 & Yes & \cite{Chowell1} \\
 Sao Paulo, Brazil & 4.6 & 2.7 & Yes & \cite{Massad1} \\
  cities in Europe and America & 4.0 & 1.2-3.0 & No & \cite{Vynnycky2007} \\
 83 cities in the UK & 3.2, 2.6 & 1.7-2.0 & No & \cite{Ferguson2, Viboud2} \\
45 cities in the USA & 2.9 & 1.7 & No & \cite{Wallinga2}\\
  RAF camp in the UK  & 2.3 & 2.9 & No & \cite{Mathews1} \\
 Featherston Military & 1.6 & 3.1 & Yes & \cite{Sertsou1} \\
 Camp, New Zealand$^c$ & 1.1 & 1.8 & Yes & \cite{Sertsou1} \\
  & 0.9 & 1.3 & Yes & \cite{Sertsou1} \\
 \hline
\end{tabular}\\
$^a$Sensitivity of the $R$ estimates to different assumptions for the serial interval was examined; $^b$Three pandemic waves were simultaneously fitted; $^c$The epidemic was observed in a community with closed contact (i.e. military camp).
\label{TableWaves}
\end{table}

\newpage

\end{document}